# Varying the VaR for Unconditional and Conditional Environments

JOHN COTTER[a]
University College Dublin

**Abstract**
Accurate forecasting of risk is the key to successful risk management techniques. Using the largest stock index futures from twelve European bourses, this paper presents VaR measures based on their unconditional and conditional distributions for single and multi-period settings. These measures underpinned by extreme value theory are statistically robust explicitly allowing for fat-tailed densities. Conditional tail estimates are obtained by adjusting the unconditional extreme value procedure with GARCH filtered returns. The conditional modelling results in iid returns allowing for the use of a simple and efficient multi-period extreme value scaling law. The paper examines the properties of these distinct conditional and unconditional trading models. The paper finds that the biases inherent in unconditional single and multi-period estimates assuming normality extend to the conditional setting.

JEL classification: G1; G10.

Keywords: Value at Risk; extreme value theory; GARCH filter; conditional risk.

[a] **Address for Correspondence:**
Dr. John Cotter,
Centre for Financial Markets,
Graduate School of Business,
University College Dublin,
Blackrock,
Co. Dublin,
Ireland.
E-mail. john.cotter@ucd.ie
Ph. +353 1 716 8900
Fax. +353 1 283 5482

# Varying the VaR for Unconditional and Conditional Environments

## 1. INTRODUCTION

Accurate forecasting of volatility is the key to successful risk management techniques. Many financial disasters attributable to failures of risk management procedures has led to greater regulatory control such as the Basle requirements, and the subject of this paper, greater emphasis on accurate modelling of market risk. Measures such as Value at Risk (VaR) incorporating new modelling procedures have been developed. This paper presents a novel procedure for scaling relatively high frequency VaR estimates encompassing the conditional distribution of price changes for the largest European stock index futures.

It is now clear that given the unconditional fat-tailed characteristic of futures price changes, the assumption of modelling market risk with the thin-tailed Gaussian distribution is inappropriate. Risk measures are misspecified with an underestimation (overestimation) bias for single-period (multi-period) settings leading to invalid risk management practices due to inappropriate capital reserves. In contrast, this paper uses extreme value theory to provide downside risk management techniques and this approach dominates other techniques for low probability and quantile combinations (Danielsson and de Vries, 2000).[1] Comparisons are made with gaussian estimates for conditional single-period and scaled multi-period intervals.

Previous applications of extreme value theory in risk management (Pownall and Koedijk, 1999; Longin, 2000; and Danielsson and de Vries, 2000) present unconditional estimates for a single-period setting. These inform risk managers of constant time-invariant VaRs averaged over the whole period of analysis thereby allowing them to assess the extent of the overall risk inherent in an asset. This paper extends the analysis twofold by presenting conditional risk estimates and scaling these for multi-periods.[2] Conditional estimates provide a profile of time-varying VaRs updated by current volatility. Conditional volatility modelling is important in many situations, for example in a short holding period during times of high volatility when there is a need to incorporate time-varying volatility signals into the investor's trading strategy. However, exclusive reliance on conditional risk measures is not a panacea for the risk manager's problems. Trading strategies would have to be continuously updated given new volatility estimates with associated high transaction costs. Thus it is also important to provide an overview of risk facing investors over long periods using unconditional measures. Thus this paper presents distinct VaR measures dealing separately with the unconditional and conditional distributions for single-period and multi-period settings providing investors with different and often diverging information.

---
[1] The relative strengths and weaknesses in applying this approach are outlined in Diebold et al. (1998).

The conditional environment is modelled with a GARCH process induced with fat-tailed characteristics through assuming student-t innovations. This results in identically and independently distributed (iid) filtered returns, and predictors of conditional returns and volatility through iteration. Investors now have risk measures conditional on the present risk and return environment facing the asset. Simulation results demonstrate the statistical properties of the time-varying procedure. Moreover, the multi-period estimates exploit the $\alpha$-root scaling law applicable in extreme value theory that only requires an iid variable. The scaling procedure advantageously requires no further estimation of any additional parameters and obtains efficiency in the scaling operation by using the highest frequency realisations. Scaling from high to low frequency has many applications, and most importantly gives investors risk scenarios for different holding periods.[3]

---

[2] See Jansen et al. (2000) for an application of scaling in an unconditional context.

The paper proceeds in the following section with an outline of extreme value theory. Section III details the methods applied to generate the conditional and unconditional risk management estimates and in particular, detailing the use of the GARCH filter and the scaling procedure for multi-periods. The statistical properties of the conditional approach are investigated through simulation. The results are presented for leading stock index futures from European bourses in section IV. Here, a description of the conditional environment, tail estimates, single and multi-period estimates and a comparison of conditional approaches are given. Finally, concluding comments are documented in section V.

**2. EXTREME VALUE THEORY**

Extreme value theory underpins the risk measures relying on order statistics where a set of logarithmic futures returns $\{R_1, R_2,..., R_n\}$ associated with days 1, 2.. n, are assumed to be independent and identically distributed (iid), and belonging to the true unknown distribution $F$.[4] We examine the maxima ($M_n$) of a sequence of n random variables

$M_n = \text{Max}\{ R_1, R_2,..., R_n\}$ (1)

---

[3] Investors also use scaling to meet regulatory requirements, for example, Basle's 10-day VaR.
[4] Much of the theory is previously documented and this will present only the salient features relevant for this study. For a comprehensive discussion of extreme value results under a wide range of distributional situations see Leadbetter et al. (1983).

The corresponding density function of $M_n$ is got from the cumulative probability relationship:

$$P\{M_n \leq r\} = P\{R_1 \leq r, \ldots, R_n \leq r\} = F(r) = 1 - ar^{-\alpha} \qquad -\infty < r < \infty \qquad (2)$$

where the scaling constant is given by a and $\alpha$ is the tail index, for $\alpha > 0$. The random variables of interest in this analysis are tail values, for example, the VaR measures the amount of possible loss exposure upto the extreme return, r.

The Fisher-Tippett theorem gives the asymptotic behaviour of the distribution detailing three types of limit laws:

Type I (Gumbell): $\Lambda(r) = \exp(-e^{-r})$ $\qquad -\infty < r < \infty$

Type II (Fréchet): $\Phi_\alpha(r) = 0$ $\qquad r \leq 0$
$\qquad\qquad\qquad\qquad\; = \exp(-r)^{(-\alpha)}$ $\qquad r > 0$

Type III (Weibull): $\psi_\alpha(r) = \exp(-(-r)^{(\alpha)})$ $\qquad r \leq 0$
$\qquad\qquad\qquad\qquad\; = 1$ $\qquad r > 0 \qquad (3)$

and for $\alpha > 0$.

The types of limit distribution are distinguished by the shape parameter $\alpha$, the tail index, in (3), detailing the asymptotic convergence rate. Of importance to this study is the Type II process that exhibits the fat-tailed characteristic in line with financial returns. This type of extreme value distribution exhibits a tail with a power decline causing a relatively slow decay for convergence towards the limit, vis-à-vis the exponential decline of the type I process. Also, the relatively slow decline in the tails generates moments that are not necessarily always finite with bounded moments upto the tail index, $\alpha$.

Fortunately, the necessary and sufficient conditions for asymptotic convergence on the type II distribution can be met using Gnedenko's theorem:

Type II (Fréchet): $\quad \lim_{t \to +\infty} \dfrac{1 - F(tr)}{1 - F(t)} = r^{-\alpha} = r^{-(1/\gamma)}$ \quad (4)

For $r > 0$, $\alpha > 0$.

This condition allows for unbounded moments and represents a tail having a regular variation at infinity property and behaves like the fat-tailed pareto distribution (Feller, 1972). By l'Hopital's rule a number of other distributions exhibit this unifying regular variation property and are fat-tailed including the independent student-t distribution and dependent ARMA process with stable innovations for $\alpha < 2$. More commonly, the finance literature models derivative first differences with second moment dependence using ARCH related specifications (see Hull and White, 1998 for an example). These processes are also unconditionally fat-tailed and display regular varying property even if the conditional distribution is thin-tailed including iid normal innovations in the case of ARCH (p) and GARCH (p, q) processes, for example ARCH (1) and GARCH (1, 1) models, although this does not apply to stochastic volatility models (de Haan et al., 1989). Furthermore assuming, for example, conditional student-t innovations, many of these processes have an unconditional distribution exhibiting even fatter tails better matching the empirical features of the financial time series.

The extension of extreme value theory in the strict iid case to the assumption of strict stationarity is fully discussed in Leadbetter et al. (1983). Support for the stationary series being an associated iid series implies that both series have the same qualitative limiting behaviour. Two conditions are required, a distributional mixing condition indicating weak long-range dependence, supported for financial data, and an anti-clustering condition rejected in the presence of ARCH type effects. Whether the extreme values of financial returns exhibit clustering is debatable (Danielsson and de Vries, 2000) but their existence would bias the estimation of the tail index (see de Haan et al., 1989). This paper adopts Huisman's et al. (2001) modified small sample tail estimator that indicates little bias for clustered data based on a simulation study of a GARCH (1, 1) process.

However, for the conditional modelling, the conditions of stationary modelling with GARCH processes are not always met (Ghose and Kroner, 1995). For stationary processes, the tail index must be greater than 2 because second unconditional moments exist requiring the sum of the GARCH parameters be less than one. This implies that certain GARCH processes, for example the IGARCH process, are inappropriate as its unconditional variance is undefined. Rather a GARCH (1, 1) model with parameters summing to less than one is applicable.

## 3. METHODOLOGY

### 3.1 Risk Management Measures

We now focus on the conditional and unconditional risk measures. Using the distribution, F(r), given in (3), VaR measures can be estimated for the conditional and unconditional distributions, providing risk managers with several pieces of information for use in their strategic responses to different risk scenarios.[5] The former provides risk managers with dynamic risk information on prospective losses occurring in a time-varying fashion whereas the latter details constant large-scale losses over long periods of analysis. From a number of stochastic processes that incorporate time-varying volatility, parametric estimation of an AR (1)-GARCH (1, 1) filter and student-t innovations with 4 degrees of freedom is used in this paper to profile the conditional distribution.[6] The statistical properties of this conditional approach and the related measures are investigated through simulation analysis discussed shortly.

---

[5] As well as quantile estimates, associated probability estimates can be obtained but are not presented for conciseness. Results available on request.

The unconditional VaR measure corresponding to a prescribed quantile, $r_p$, from the tail of the marginal distribution, $F(r)$ is:

VaR $[Rr_p] =  r_{m, n} (m/np)^\gamma$ (5)

Where $\gamma$ is the Hill (1975) semi-parametric tail estimator.[7]

In terms of robustness, the Hill estimator is recognised as the most efficient semi-parametric tail estimator (Kearns and Pagan, 1997), and also, it operates analogously with extreme value theory by dealing with order statistics. This downside risk estimator has the same statistical properties as the quantile measure:

$\gamma = 1/\alpha = (1/m - 1) \sum [\log (-r_i) - \log (-r_m)]$    for i = 1,...., m - 1. (6)

The Hill estimator is asymptotically normal, $(\gamma - E\{\gamma\})/(m)^{1/2} \approx (0, \gamma^2)$ (Hall, 1982). Choice of the optimal threshold value, m, is nontrivial in tail estimation although a modified estimator removing bias in small samples is available (Huisman et al., 2001).

Second, we focus on the conditional risk estimates. The unconditional VaR estimates are modified according to the conditional distribution generating a set of predictive time-varying estimates. For example, the conditional VaR estimate for a one-day holding period is obtained with:

VaR$[R^t r_p] =  \mu_{t+1} + \sigma_{t+1}$VaR $[Zr_p]$ (7)

---

[6] The choice of the conditional distribution is due to the GARCH process with student-t innovations satisfying the fat-tailed regular variation at infinity property.
[7] VaR measures are not without criticism as they do not examine losses beyond the chosen quantile

The series Z represents the returns series filtered by the GARCH estimate of conditional standard deviation, that also predicts conditional mean and volatility through iteration. Appropriate filtering of this type can result in iid variables (see Taylor, 1986; and Andersen et al., 2000, for applications with different methods). The application of a simple and efficient α-root scaling law applicable for extreme values allows for the extension to multi-period risk management forecasts.

Thus far the discussion is for a single-period. For multi-period forecasts the risk measures are scaled by a α-root of time multiplication factor. Illustrating the scaling law for the conditional measures, taking two return sequences, the single-period $R_t$, and the multi-period $M[R_t]$ sum of n periods single returns $M[R_t] = \sum_{t=1}^{n} R_t$, we can adjust the asymptotic distribution of the fat-tailed Fréchet distribution in (3) by applying Feller's theorem (Feller, 1972, VIII.8):

$$P\{\sum_{t=1}^{n} R_i \leq r = qF(r) \tag{8}$$

Asymptotically this implies that the scaling factor for the VAR estimates are easily adjusted by q (for $q = n^{1/\alpha}$). This law is strictly applicable for an iid variable and bias may occur for the multi-period forecasts of the unconditional sequences (Jansen et al., 2000). However, this bias would be reduced dramatically in the conditional multi-period application due to the near iid structure of the filtered returns.

Our forecasts extended for a multi-period setting gives a quantile estimate of

$$M[VaR[R^t r_p]] = q[VaR[R^t r_p] \tag{9}$$

---

(see Artzner et al. 1999, for a discussion of one such alternative measure).

Advantageously the extension for the extreme value estimates is similar to a gaussian distribution with its square root of time scaling factor (Diebold et al., 1998). In an unconditional setting, the scaling law application infers a reverse of the estimation dilemma in comparing gaussian and Extreme Value estimates. The single-period underestimation problem assuming normality (Cotter, 2001) reverses to become a multi-period overestimation problem when scaled upwards. This is due to the fat-tailed distribution exhibiting a finite variance ($\alpha > 2$) and resulting in $\sqrt{n} > n^{1/\alpha}$ (see Dacorogna et al., 1995; for further details). The paper uses a small simulation to explore the comparison of the scaling laws for a conditional setting.

Advantageously, extreme value scaling for multi-period forecasts can be completed without re-estimation of any additional parameters. More importantly, the tail index estimates, $1/\alpha$, are most efficient at highest frequencies due to their fractal nature (Dacorogna et al., 1995). The increased efficiency for high frequency tail estimation is due to negative sample size effects for low frequency returns. However, high frequency tail estimation empirically does involve the possibility of a downward bias (although tail estimates are theoretically invariant with respect to time aggregation).[8] Combining these outcomes, the bias gains of estimating tail values for aggregated returns are dominated by the efficiency gains in using single-period observations with their increased sample size.

### *3.2 GARCH Filtering Procedure*

In order to obtain the conditional risk estimates, related measures of the mean and variance parameters, and more importantly, the sequence Z are required. From a number of alternative processes applied in a similar fashion including ARMA-GARCH (Barone-Adesi et al., 1999) and AR-GARCH (McNeil and Frey, 2000) specifications, the latter is chosen as it fulfils the objectives of obtaining a forecast of the conditional expected value $\mu_{t+1}$ through the AR component of the filter, the conditional variance $\sigma_{t+1}$ and residual sequence Z through the GARCH component. Assuming that a sequence of returns, R, is related to Z by:

$$R_t = \mu_t + \sigma_t Z_t \tag{10}$$

---

[8] This downward bias for relatively small samples can be circumvented with the appropriate tail estimation procedure (Huisman et al., 2001).

Conditional on the information upto day t. The sequence, Z, introduces randomness by being a (near) iid sequence and is by the extreme value methods applying the Hill estimator, and is used to calculate the conditional quantile estimator. Extensions for multi-period conditional forecasts are obtained from the extreme value α-root scaling law.

The main assumption of GARCH models is that the conditional second moment, $\sigma_t$, has a degree of persistence focusing on volatility clustering with periods of high (low) volatility followed by similar periods of high (low) volatility. Our conditional VaR measure explicitly adapts the risk values for this feature by assuming that volatility is time-varying. Furthermore, the volatility clustering feature in GARCH (1, 1) models gives rise to fat-tails due to positive excess kurtosis assuming the unconditional fourth moment exists. Formally an ARMA process defines the volatility term:

$$\sigma^2_t = \alpha_0 + \alpha_1 R_{t-1}^2 + \beta_1 \sigma^2_{t-1} \tag{11}$$

for $\alpha_0$, $\alpha_1$, and $\beta > 0$; and $0 < \alpha_1 + \beta_1 < 1$ to fulfil the strictly stationary requirement in line with the unconditional extreme value framework. β measures the persistence in volatility. To formally account for fat-tails the GARCH model is fitted to the data assuming the conditional density follows a student-t distribution with 4 degrees of freedom.[9] Specifically Mikosch and Starcia (2000) show that under the assumption of a student t conditional distribution, the GARCH (1, 1) process belongs to the marginal distribution F

$$\int_{-\infty}^{\infty} \ln\left|\alpha_1 z^2 + \beta_1\right| g(z)\,dz < 0, \qquad \alpha_0 > 0 \tag{12}$$

where g(z) is a student-t density with 4 degrees of freedom and the prerequisite is easily verified numerically.

*3.3 Monte Carlo Simulation:*

Here we examine the statistical properties of the proposed conditional estimators.[10] The quantile estimator is obtained in the time-varying environment by a stochastic process represented by a GARCH (1, 1) model. Specifically a stationary process is assumed by having $\alpha_0 > 0$, and $0 < \alpha_1 + \beta_1 < 1$.[11] Parameters $\alpha_0 = 0.1$, $\alpha_1 = 0.15$, and $\beta_1 = 0.8$ are chosen. The GARCH model incorporates volatility clustering that is an important feature of financial returns series. The simulated data also meets the requirement for financial returns of having a heavy tailed conditional distribution have student-t innovations with 4 degrees of freedom. To examine the impact of scaling from single-periods to multi-periods the simulations are repeated for n = 2, 4 and 5. A sample size of 2000 is chosen with 200 replications and the average results are presented in table 1. The small sample modified Hill estimator suggested by Huisman et al. (2001) is used in the quantile estimates and then scaled using the α-

---

[9] The fitted GARCH model can assume a range of underlying distribution function including the commonly assumed thin-tailed Gaussian density as utilised by Riskmetrics in their development of VaR measures. However, this would underestimate the magnitude of tail behaviour.
[10] Monte Carlo simulations for a related unconditional approach are discussed in Jansen et al. (2000).
[11] Non-stationary series which have undefined second moments are avoided such as EGARCH with the leverage term λ = 1 and IGARCH processes (Ghose and Kroner, 1995).

root scaling law as outlined requiring no further estimation at lower frequencies.[12]

Quantile estimates are presented for the single-period and multi-periods chosen to demonstrate the conditional procedure. The true quantiles are in parentheses. The precision of the findings is favourable with the predicted values close to the true values. The results deteriorate slightly going from relatively high probabilities of 95% in contrast to low probabilities of 99%. The estimates also fare well for the multi-period extensions although the predicted values tend to overestimate relative to the true values. This overestimation bias tends to increase moving to longer multi-period settings. Again there is a deterioration of the results moving from high to low probabilities supporting the properties of the unconditional procedure (see Jansen et al., 2001). However, all the predicted quantities are reasonably close to the true values supporting the modeling procedure.

INSERT TABLE 1 HERE

## 4. EMPIRICAL FINDINGS

The futures analysed entail the main twelve stock index contracts traded on the respective European exchanges. Basic summary details on the contracts are given in table 2. Futures returns use the first difference of the natural logarithm of daily closing prices. The continuous time series of returns is generated for each contract using prices for the nearest maturing contract, and within this, upto the last trading day prior to the delivery month before overlapping with the next maturity. Futures returns like other speculative assets display a number of common characteristics

---

[12] This uses a weighted least squares regression of Hill estimates against associated numbers of tail estimates, $\gamma(m) = \beta_0 + \beta_1 + \varepsilon(m)$ for m = 1,....,η. The approach minimises heteroskedasticity in the regression's error term with the weighted least squares approach. Huisman et al. (2001) find that the estimator works well for small samples (similar in size to that analysed here) and with GARCH type

including weak stationarity, skewness, leptokurtosis and non-normality. The fat-tailed characteristic present for all contracts is illustrated by the Quantile-Quantile (Q-Q) plot for the Dutch AEX contract in figure 1 with both lower and upper values diverging substantially from the corresponding normal values.

INSERT TABLE 2 HERE

INSERT FIGURE 1 HERE

Maximum Likelihood estimates of the conditioning variables from fitting the AR-GARCH (1, 1) model with student-t innovations, and the dependence structure of the futures returns and filtered series are given in table 3. The conditioning mean parameter is strongest for the PSI20 contract with an AR coefficient of 0.134. The time-varying GARCH parameters are in line with financial studies measured at daily intervals indicating that past volatility impacts current volatility. Also the parameters indicate strict stationarity with the summation of the GARCH coefficients being less than one. Persistence of past squared returns and volatility is strongest for the Portuguese and UK futures respectively and investors should be aware of dependency in their investment strategies where volatility is not constant. Analysis of the Ljung-Box statistics confirms this strong serial dependence for the returns series. The conditional specification appears well specified however, with negligible serial correlation resulting in near iid filtered returns series. Advantageously, the extreme value scaling law for the multi-period setting is employed for these iid variables.

INSERT TABLE 3 HERE

---

dependency. The associated number of tail estimates, $m_{hkkp}$, is extrapolated based on the modified Hill estimator.

The quantile statements are derived from the tail estimates, and values for the semi-parametric Hill estimator are given in table 4 for the returns and filtered series. Importantly, a lack of stability in the Hill estimates can affect the VaR measures and their qualitative inference. The extent of the problem is such that a 'Hill horror plot' illustrating the variability of tail index estimates for different thresholds is beneficial (Embrechts et al., 1997). Given the possible variability in tail estimates, a pragmatic approach is adopted in the development of risk management measures by combining previously supported techniques. First, it follows Phillips et al. (1996), and calculates an optimal threshold value for each contract based on a bootstrap procedure of $m = M_n = \{\lambda n^{2/3}\}$ where $\lambda$ is estimated adaptively by $\lambda = |\gamma_1/2^{1/2}(n/m_2(\gamma_1 - \gamma_2)|^{2/3}$. Second it takes account of small sample bias and the impact of tail clusters by using the modified Hill estimator, $\gamma_{hkkp}$, proposed by Huisman et al. (2001). Finally a related qualitative approach that ensures that Hill estimates do not suffer from instability is obtained from a Hill plot.

INSERT TABLE 4 HERE

The importance of appropriate tail estimation procedures is evident from comparing methods. Huisman et al. (2001) find that the small sample bias results in lower tail estimates and this is removed by their modified estimator as can be seen by a comparison with the Phillips et al. (1996) estimates. The Hill estimator used in the quantile estimates remain stable as demonstrated by the Hill plot for the IBEX index in figure 2 detailing stable estimates over a range of threshold values. Concentrating on the Huisman et al. (2001) estimates the modified small sample values range between two and four, verifying previous studies on financial returns (Loretan and Phillips, 1994). The existence of a finite second moment is supported using a difference in means statistic giving credence to the use of stationary GARCH models and appropriateness of the α-root scaling law. An inverse relationship exists between tail estimates and the degree of tail fatness suggesting that the Portuguese PSI20 contract exhibits greater potential for more extreme returns for prospective investors. Unfortunately it is in times of financial disasters that the role of tail fatness and the associated extreme returns become paramount. Here risk management procedures face possible crises and accurate modelling of these events is vital.

INSERT FIGURE 2 HERE

A spectrum of single-period, and multi-period using the α-root scaling law, unconditional VaR estimates is presented in table 5. These estimates suggest that, for example, there is a 95% probability that the loss on the BEL20 contract is less than or equal to 1.45%, with the MIF contract being most risky. For the 99.5% level the Portuguese PSI20 contract is now most risky with a VaR of 6.32%. The multi-period quantiles for blocks of trading days relying only on the relatively high frequency daily data are presented in the last six columns. Dealing with the lower probability level, the VaR for the BEL20 contract is 5.30% for weekly intervals with an expected occurrence of once every 200 weeks (1/1 – p) on average, in comparison to 12.65% for the most risky PSI20 futures with the same occurrence ratio. The Swiss contract has the lowest potential for exteme losses. These multi-period estimates do not require further estimation and benefit from measurement efficiency by using the high frequency daily data to measure the tail estimator.

INSERT TABLE 5 HERE

Switching our attention to the conditional risk management measures the single-period and multi-period forecasted VaR estimates are given in table 6. As all returns are calculated upto February 28 1999, the forecasts are for single-periods and multi-periods beginning 1 March 1999.[13] The dynamic estimates rely on the filtered series, Z, which is a (near) iid series, and conditional mean and volatility values obtained through iteration of the AR(1)-GARCH (1, 1) process. The VaR estimates reach as high as 12.15% for the PSI20 contract with 99.5% confidence for weekly periods whereas the lowest comparable estimate occurs for the BEL20 contract. In general most of the contracts exhibit higher conditional daily VaRs indicating higher volatility during the February 1999.

INSERT TABLE 6 HERE

There are clearly diverging results between conditional and unconditional environments with an increase (decrease) in tail risk occurring for the PSI20 (BEL20) futures in the time-varying context. These conditional measures are weighted towards the trading environment leading upto March 1 and any volatility considerations affecting specific contracts at that time would affect the estimated dynamic risk measures. Thus investors should be aware of the important impact of the time dependent conditional environment in their development of risk management procedures to avoid disasters associated with extreme returns.

These multi-period conditional forecasts scale the single-period estimates, and as noted, this scaling law is strictly applicable for an iid process as evidenced by the dependence structure of the filtered returns sequences in table 2. This extends the single-period case advocated by McNeil and Frey (2000) by exploiting Feller's theorem to apply the α-root scaling law ($n^{1/\alpha}$). Also, the multi-period forecasts have the advantage that the conditional environment is not measured at lower frequencies thereby avoiding losing the unique stylised features of relatively high frequency realisations and avoiding the dampening of volatility estimates.

---

[13] The forecasts for the Danish KFX and Swiss contracts are exceptions to this, dealing with 19 December 1998 and 1 July 1999 respectively.

Previous studies have compared methods over single-periods for the unconditional setting and found that the much-used normal distribution underestimates tail behaviour and associated risk measures relative to extreme value theory. Furthermore, these results reverse over multi-periods where the gaussian $\sqrt{n}$ scaling law overestimates VaR estimates relative to the α-root scaling law (Dacorogna et al., 1995). The use of GARCH models assuming a gaussian underlying distribution is extremely popular in the finance literature and industry, and quantile measures are presented in table 6 from fitting a GARCH (1, 1) specification.[14]

These estimates allow for comparison between gaussian and extreme value risk measures. First, the unconditional underestimation assuming normality is substantiated for the conditional VaRs demonstrating the impact of the unconditional assumptions of the GARCH model. Furthermore, this underestimation reverses to become an overestimation when scaling upwards to multi-periods, for example, the 95% VaR for the Danish KFX contract over 2 days increases to 2.54% (2.63%) for extreme value (gaussian) estimates.[15] Thus as the futures data exhibit finite variance, the expansion of the $\sqrt{n}$ scaling law for normality exceeds that of the extreme value $\alpha$-root scaling law even in a time-varying setting. On moving to larger intervals the divergence between gaussian and extreme value estimates increases further with the 95% VaR for the KFX contract now becoming 3.29% and 4.16% respectively. Risk management procedures should be adjusted given these results. Namely, conditional estimates assuming normality underestimates potential risk and should incorporate an additional risk element, whereas in contrast, the scaling procedure of normality overestimates the impact of conditional attributes leading to trading practices that are too conservative.

---

[14] For example, the widely applied RiskMetrics™ method is a special case of such a GARCH model. Here they assume the unconditional environment is gaussian and they specify the persistence parameters of 0.94 for past volatility with the ARCH coefficient equal to 0.06. This results in the IGARCH specification.

**5 SUMMARY AND CONCLUSION**

This paper presents market risk measures accounting for the fat-tailed characteristic of futures returns. Extreme value methods based on order statistics model the tail values of a distribution in an unconditional and conditional setting. The conditional risk estimates rely on using an AR(1)-GARCH (1, 1) filter. These conditional measures profit from applying the parametric GARCH model leaving near iid filtered returns allowing for the $\alpha$-root scaling law giving multi-period estimates. Scaling allows risk managers to assess likelihood of losses across intervals estimated parsimoniously and efficiently. An application for a range of European stock index futures is given so investors can infer risk patterns across markets and develop trading strategies according to their risk preferences.

---

[15] Not all contracts exhibit larger Gaussian estimate for this multi-period setting, for example the BEL20 contract, however there is a reversal for this contract in extending to larger periods, and in general, this reversal should take place on moving to some multi-period interval.

The risk management process is aided considerably by the procedures detailed in this paper. For instance, as accurate risk measures rely on appropriate modelling it is important to use the modified small sample Hill estimator minimising overestimation of tail fatness. Furthermore, the impact of the conditional distribution on risk estimates is explored. Investors can now update their decision making process by focusing on volatility levels for the current period by applying the most recent price information. These volatility levels are time-varying and periods of high (low) volatility result in an increase (decrease) in unconditional VaR estimates. This paper finds that the Portuguese futures exhibit the most volatile market environment resulting in the highest conditional VaR estimates. In general the futures exhibit diverging potential for extreme returns in comparing the conditional and unconditional estimates providing risk managers with distinct risk profiles.

These conditional measures are easily scaled for lower frequencies using the α-root scaling law giving the most efficient multi-period risk estimates. By way of comparison scaled conditional estimates assuming the underlying distribution is gaussian are computed. The results confirm the underestimation of risk estimates for a single-period setting albeit for the conditional environment. Furthermore, for multi-period settings the $\sqrt{n}$ scaling law results in overestimation of the conditional estimates relative to the extreme value α-root scaling law. Investors should be aware that the normality estimates result in too conservative risk estimates even in a conditional environment and adjust their capital reserves accordingly.

ACKNOWLEDGEMENTS: The author would like to thank participants at the State of the Art on Value at Risk Conference, Forecasting Financial Markets Annual Conference, European Financial Management Association Annual Conference, Global Association of Risk Professionals Annual Conference, Irish Economic

Association's Annual Conference, Statistical and Social Inquiry Society of Ireland Seminar, Erasmus University Rotterdam Seminar, Queen's University Seminar, Casper de Vries, Kees Koedijk, Francois Longin, Donal McKillop, Ronan O' Connor and Franz Palm for their helpful comments on this paper. University College Dublin's President's Research Awards and a Faculty grant provided financial support.

**Figure 1**

Q-Q Plot of AEX Futures Contract

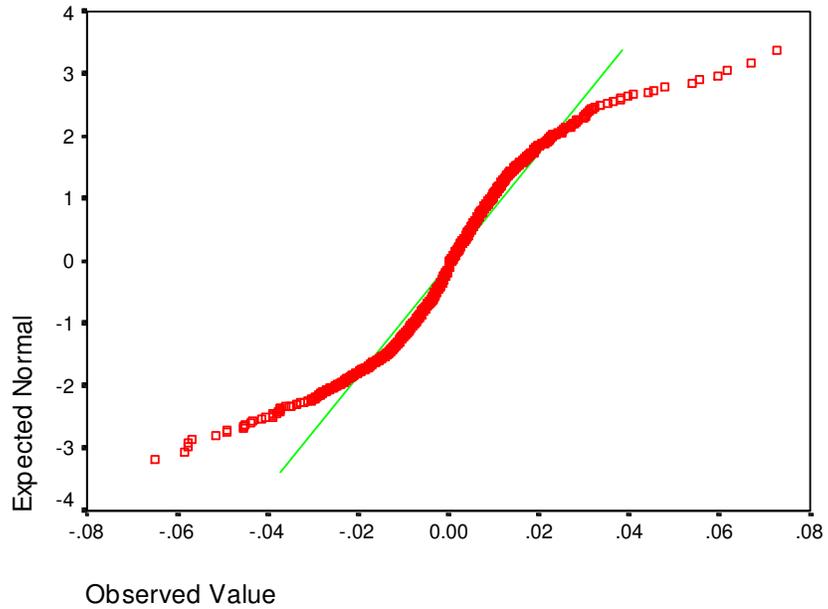

This figure plots the quantile of the empirical distribution of the AEX futures index returns against the normal distribution. The straight line represents a gaussian quantile plot whereas the curved line represents the quantile plot of the empirical distribution of the AEX contract. The extent to which these AEX returns diverge from the straight line indicates the fat-tail characteristic.

**Figure 2**

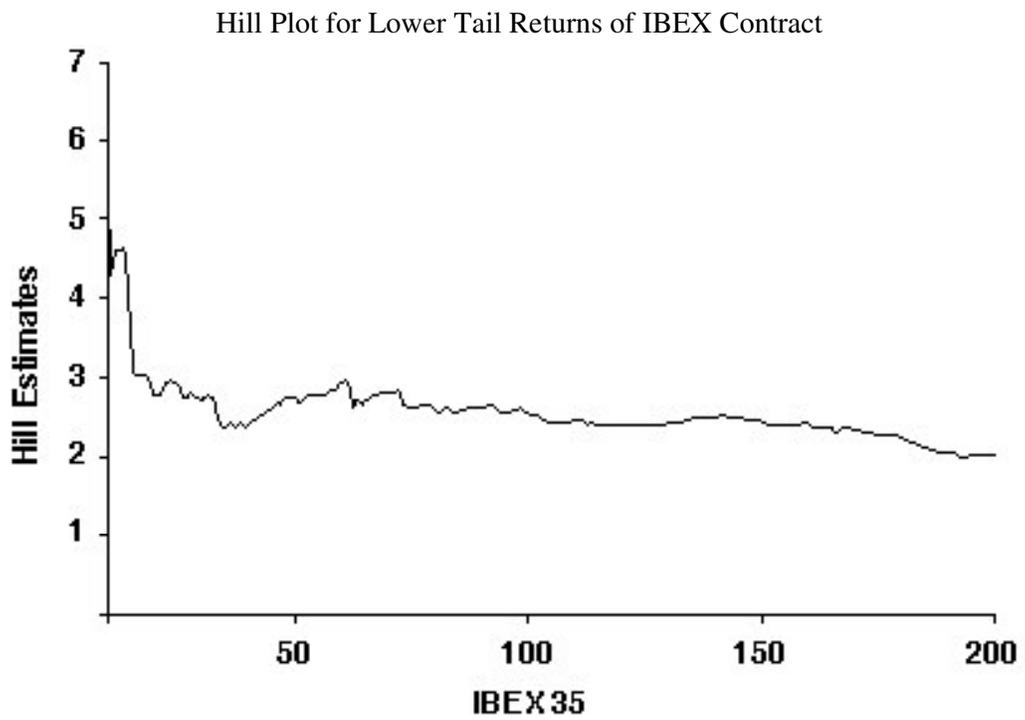

Hill Plot for Lower Tail Returns of IBEX Contract

The figure represents a Hill plot for various thresholds, m, determining whether there is stability of the Hill estimates.

**Table 1**
Simulated GARCH (1, 1) with Student-t innovations and Scaling Procedure

|  | Single-period | Multi-period | | |
| --- | --- | --- | --- | --- |
|  | n = 1 | n = 2 | n = 4 | n = 5 |
| **Quantile** | | | | |
| $r_p 95$ | 7.0413 | 9.1925 | 12.0010 | 13.0764 |
|  | (7.0900) | (8.4315) | (10.0268) | (10.6020) |
| $r_p 99$ | 13.0764 | 17.0714 | 22.2869 | 24.2842 |
|  | (13.6000) | (16.1732) | (19.2333) | (20.3367) |

The values in this table represent averages of 200 replications from a sample size of 2000. The blocks for the multi-periods used are n = 2, n = 4 and n = 5 corresponding to 2 days, 4 days and weekly intervals. The quantile estimates are based on Huisman et al. (2001) tail estimates for the simulated data. The theoretical quantiles are in parentheses.

**Table 2**

Summary Details of European Stock Index Futures Analysed

| Contract | Country | Period | Unit Root | Skewness | Kurtosis | Normality |
|---|---|---|---|---|---|---|
| BEL20 | Belgium | 01/12/1992 | -1156.75 | -0.11† | 4.4 | 0.07 |
| KFX | Denmark | 01/06/1992 | -1648.93 | -0.33 | 5.19 | 0.08 |
| CAC40 | France | 01/12/1988 | -2466.2 | -0.08† | 3.36 | 0.05 |
| AEX | Holland | 01/12/1990 | -2631.2 | -0.33 | 5.94 | 0.08 |
| DAX | Germany | 01/12/1988 | -1973.46 | -0.56 | 8.31 | 0.08 |
| MIF30 | Italy | 01/12/1994 | -1142.23 | -0.06† | 2.27 | 0.06 |
| OBX | Norway | 01/12/1992 | -1550.45 | 0.32 | 97.69 | 0.18 |
| PSI20 | Portugal | 01/10/1996 | -546.65 | -0.87 | 7.95 | 0.11 |
| IBEX35 | Spain | 01/06/1992 | -1634.7 | -0.49 | 4.72 | 0.07 |
| OMX | Sweden | 01/03/1990 | -2364.04 | -0.27 | 8.93 | 0.07 |
| FTSE100 | UK | 01/06/1984 | -2321.14 | -1.18 | 18.78 | 0.05 |
| SWISS | Switzerland | 01/12/1990 | -1709.49 | -0.5 | 9.3 | 0.06 |

Datastream provided the data. The period gives the respective starting dates for each contract and ends on 23/2/99 for all except KFX (18/12/1998) and Switzerland (30/06/1999). The Phillips Perron test examines for unit roots. Normality is examined with the Kolmogorov-Smirnov test. † represent insignificant at the five percent level.

**Table 3**

Conditional Modelling of European Stock Index Futures

| Contract | AR | $\alpha_0$ | $\alpha_1$ | $\beta_1$ | R(12) | $R^2$(12) | Z(12) | $Z^2$(12) |
|---|---|---|---|---|---|---|---|---|
| BEL20 | 0.077 | 9.46E-07 | 0.049 | 0.915 | 41.488 | 321.104 | 12.720 | 14.350 |
|  | (0.027) | (3.96E-07) | (0.011) | (0.018) | 0.000 | 0.000 | [0.390] | [0.279] |
| KFX | 0.049 | 1.73E-06 | 0.064 | 0.876 | 12.110 | 318.665 | 14.310 | 9.214 |
|  | (0.024) | (5.55E-07) | (0.011) | (0.020) | 0.437 | 0.000 | [0.281] | [0.685] |
| CAC40 | 0.006 | 3.14E-06 | 0.050 | 0.902 | 27.361 | 348.855 | 19.950 | 3.434 |
|  | (0.020) | (8.90E-07) | (0.008) | (0.015) | 0.007 | 0.000 | [0.068] | [0.992] |
| AEX | -0.050 | 1.03E-06 | 0.046 | 0.918 | 36.448 | 258.129 | 13.020 | 0.493 |
|  | (0.022) | (3.72E-07) | (0.008) | (0.013) | 0.000 | 0.000 | [0.368] | [1.000] |
| DAX | -0.008 | 1.45E-06 | 0.060 | 0.889 | 40.749 | 961.226 | 19.130 | 5.304 |
|  | (0.020) | (3.68E-07) | (0.009) | (0.014) | 0.000 | 0.000 | [0.085] | [0.947] |
| MIF30 | -0.034 | 5.27E-06 | 0.071 | 0.875 | 32.027 | 445.698 | 14.450 | 8.521 |
|  | (0.032) | (2.25E-06) | (0.016) | (0.027) | 0.001 | 0.000 | [0.273] | [0.743] |
| OBX | 0.067 | -1.74E-16 | 0.094 | 0.600 | 11.564 | 45.477 | 0.009 | 0.035 |
|  | (0.020) | (9.74E-16) | (0.003) | (0.035) | 0.481 | 0.000 | [1.000] | [1.000] |
| PSI20 | 0.134 | 1.68E-06 | 0.120 | 0.781 | 30.401 | 187.883 | 8.310 | 6.213 |
|  | (0.040) | (9.10E-07) | (0.028) | (0.038) | 0.002 | 0.000 | [0.761] | [0.905] |
| IBEX35 | -0.007 | 4.38E-06 | 0.050 | 0.891 | 25.791 | 423.089 | 18.030 | 13.710 |
|  | (0.024) | (1.53E-06) | (0.010) | (0.022) | 0.011 | 0.000 | [0.115] | [0.320] |
| OMX | 0.006 | 4.90E-06 | 0.060 | 0.876 | 16.061 | 324.787 | 8.716 | 35.330 |
|  | (0.021) | (1.13E-06) | (0.010) | (0.017) | 0.188 | 0.000 | [0.727] | [0.000] |
| FTSE100 | 0.012 | 1.25E-06 | 0.043 | 0.928 | 20.213 | 49.866 | 15.100 | 2.353 |
|  | (0.016) | (3.11E-07) | (0.006) | (0.009) | 0.063 | 0.000 | [0.236] | [0.999] |
| SWISS | -0.033 | 4.53E-06 | 0.060 | 0.806 | 25.160 | 1065.037 | 18.040 | 1.342 |
|  | (0.025) | (1.34E-06) | (0.014) | (0.042) | 0.014 | 0.000 | [0.114] | [1.000] |

The AR-GARCH specification assumes student-t innovations with 4 degrees of freedom. Marginal significance levels using Bollerslev-Wooldridge standard errors are displayed by parentheses. Ljung-Box test are for the returns (R) and filtered (Z) series. Ljung-Box test are for the squared returns ($R^2$) and squared filtered ($Z^2$) series. Marginal significance levels for the Ljung-Box tests given in brackets. * denotes significance at the 5% level.

**Table 4**

Downside Tail Estimates for Stock Index Futures

| Contract | Returns | | | | Filtered Returns | | | |
|---|---|---|---|---|---|---|---|---|
| | $m_p$ | $\gamma_p$ | $m_{hkkp}$ | $\gamma_{hkkp}$ | $m_p$ | $\gamma_p$ | $m_{hkkp}$ | $\gamma_{hkkp}$ |
| **BEL20** | 63 | 2.81 | 45 | 3.02 | 68 | 3.30 | 51 | 3.98 |
| | | (0.35) | | (0.45) | | (0.40) | | (0.56) |
| **KFX** | 72 | 2.65 | 40 | 2.86 | 80 | 2.88 | 36 | 3.56 |
| | | (0.31) | | (0.45) | | (0.32) | | (0.59) |
| **CAC40** | 100 | 2.97 | 89 | 3.25 | 102 | 3.95 | 117 | 3.51 |
| | | (0.3) | | (0.34) | | (0.39) | | (0.32) |
| **AEX** | 93 | 3.04 | 64 | 3.24 | 91 | 3.26 | 85 | 3.15 |
| | | (0.32) | | (0.41) | | (0.34) | | (0.34) |
| **DAX** | 109 | 2.93 | 65 | 3.05 | 100 | 4.01 | 105 | 4.1 |
| | | (0.28) | | (0.38) | | (0.4) | | (0.40) |
| **MIF30** | 55 | 3.31 | 25 | 3.3 | 59 | 3.54 | 53 | 4.06 |
| | | (0.45) | | (0.66) | | (0.46) | | (0.56) |
| **OBX** | 71 | 2.04 | 22 | 2.45 | 69 | 3.23 | 40 | 3.14 |
| | | (0.24) | | (0.52) | | (0.39) | | (0.50) |
| **PSI20** | 41 | 1.91 | 18 | 2.32 | 43 | 2.41 | 21 | 3.51 |
| | | (0.3) | | (0.55) | | (0.37) | | (0.77) |
| **IBEX35** | 74 | 2.62 | 25 | 2.92 | 80 | 3.41 | 109 | 3.26 |
| | | (0.3) | | (0.58) | | (0.38) | | (0.31) |
| **OMX** | 88 | 2.59 | 17 | 2.85 | 96 | 3.46 | 85 | 3.4 |
| | | (0.28) | | (0.69) | | (0.35) | | (0.37) |
| **FTSE100** | 126 | 2.99 | 127 | 3.00 | 126 | 4.22 | 211 | 3.75 |
| | | (0.27) | | (0.27) | | (0.38) | | (0.26) |
| **SWISS** | 72 | 2.81 | 61 | 3.02 | 81 | 2.96 | 46 | 3.09 |
| | | (0.33) | | (0.39) | | (0.33) | | (0.46) |

Hill tail estimates, $\gamma$, are calculated for each futures index returns, and filtered returns series using the AR(1)-GARCH(1, 1) model detailed in text. The number of values in the respective tails, $m_p$, and the associated Hill estimates, $\gamma_p$, follows Phillips et al. (1996). The number of values in the respective tails, $m_{hkkp}$, and the associated Hill estimates, $\gamma_{hkkp}$, follows Huisman et al. (2001). Standard errors are presented in parenthesis for each tail value.

**Table 5**

Single-period and Multi-period Unconditional VaR Estimates for European Stock Index Futures

| | Single-period | | Multi-period | | | | | |
| --- | --- | --- | --- | --- | --- | --- | --- | --- |
| | $r_p95$ | $r_p99.5$ | $r_p95$ | | | $r_p99.5$ | | |
| Contract | n = 1 | | n = 2 | n = 4 | n = 5 | n = 2 | n = 4 | n = 5 |
| **BEL20** | 1.45 | 3.11 | 1.83 | 2.30 | 2.47 | 3.92 | 4.93 | 5.30 |
| **KFX** | 1.81 | 4.06 | 2.31 | 2.95 | 3.19 | 5.17 | 6.59 | 7.12 |
| **CAC40** | 2.45 | 4.19 | 3.04 | 3.76 | 4.02 | 5.19 | 6.42 | 6.87 |
| **AEX** | 2.01 | 4.09 | 2.49 | 3.08 | 3.30 | 5.06 | 6.27 | 6.72 |
| **DAX** | 1.75 | 3.73 | 2.20 | 2.76 | 2.97 | 4.68 | 5.88 | 6.32 |
| **MIF30** | 2.65 | 5.32 | 3.26 | 4.03 | 4.31 | 6.56 | 8.09 | 8.66 |
| **OBX** | 1.56 | 4.01 | 2.08 | 2.76 | 3.02 | 5.31 | 7.05 | 7.73 |
| **PSI20** | 2.42 | 6.32 | 3.26 | 4.39 | 4.83 | 8.52 | 11.49 | 12.65 |
| **IBEX35** | 2.36 | 5.18 | 2.99 | 3.79 | 4.09 | 6.57 | 8.33 | 9.00 |
| **OMX** | 2.55 | 5.73 | 3.26 | 4.15 | 4.49 | 7.31 | 9.32 | 10.08 |
| **FTSE100** | 1.62 | 3.50 | 2.05 | 2.58 | 2.78 | 4.41 | 5.56 | 5.99 |
| **SWISS** | 1.39 | 2.99 | 1.75 | 2.21 | 2.37 | 3.76 | 4.73 | 5.09 |

The values in this table represent the unconditional VaR quantiles for different confidence intervals, for example $r_p95$ is the 5% level. The estimates use Hill estimators based on the Huisman et al. (2001) procedure. The blocks of returns used are two days n = 2, four days n = 4 and five days (weekly) n = 5. Values are expressed in percentages.

**Table 6**

Single-period and Multi-period Conditional VaR Estimates for European Stock Index Futures

| Contract | Single-period | | Multi-period | | | | | |
|---|---|---|---|---|---|---|---|---|
| | $r_p95$ | $r_p99.5$ | $r_p95$ | | | $r_p99.5$ | | |
| | n = 1 | | n = 2 | n = 4 | n = 5 | n = 2 | n = 4 | n = 5 |
| BEL20 | 1.81 | 3.29 | 2.16 | 2.57 | 2.72 | 3.91 | 4.65 | 4.92 |
| | (1.49) | (3.24) | (2.10) | (2.97) | (3.32) | (4.58) | (6.48) | (7.24) |
| KFX | 2.09 | 3.90 | 2.54 | 3.09 | 3.29 | 4.74 | 5.76 | 6.14 |
| | (1.86) | (3.89) | (2.63) | (3.72) | (4.16) | (5.50) | (7.77) | (8.69) |
| CAC40 | 2.41 | 4.64 | 2.94 | 3.58 | 3.81 | 5.65 | 6.88 | 7.33 |
| | (2.02) | (4.16) | (2.86) | (4.04) | (4.51) | (5.89) | (8.33) | (9.31) |
| AEX | 2.39 | 5.16 | 2.97 | 3.70 | 3.98 | 6.44 | 8.02 | 8.61 |
| | (1.92) | (3.84) | (2.71) | (3.84) | (4.29) | (5.43) | (7.69) | (8.59) |
| DAX | 2.24 | 4.01 | 2.66 | 3.14 | 3.32 | 4.75 | 5.62 | 5.94 |
| | (1.62) | (3.54) | (2.29) | (3.23) | (3.62) | (5.00) | (7.07) | (7.91) |
| MIF30 | 3.27 | 5.66 | 3.88 | 4.60 | 4.86 | 6.71 | 7.96 | 8.41 |
| | (2.78) | (5.28) | (3.94) | (5.57) | (6.23) | (7.47) | (10.57) | (11.81) |
| OBX | 1.49 | 3.75 | 1.86 | 2.33 | 2.50 | 4.69 | 5.87 | 6.31 |
| | (1.47) | (4.36) | (2.08) | (2.94) | (3.29) | (6.16) | (8.71) | (9.74) |
| PSI20 | 2.76 | 7.68 | 3.36 | 4.10 | 4.36 | 9.36 | 11.40 | 12.15 |
| | (2.47) | (6.49) | (3.49) | (4.94) | (5.52) | (9.18) | (12.99) | (14.52) |
| IBEX35 | 2.60 | 5.14 | 3.21 | 3.97 | 4.25 | 6.36 | 7.87 | 8.43 |
| | (2.34) | (5.90) | (3.31) | (4.68) | (5.23) | (8.35) | (11.80) | (13.20) |
| OMX | 2.62 | 5.06 | 3.21 | 3.93 | 4.20 | 6.21 | 7.61 | 8.13 |
| | (2.40) | (5.69) | (3.39) | (4.79) | (5.36) | (8.04) | (11.37) | (12.72) |
| FTSE100 | 2.20 | 4.03 | 2.65 | 3.19 | 3.39 | 4.85 | 5.83 | 6.19 |
| | (1.68) | (3.59) | (2.38) | (3.36) | (3.76) | (5.08) | (7.18) | (8.03) |
| SWISS | 1.78 | 3.58 | 2.23 | 2.79 | 2.99 | 4.48 | 5.60 | 6.02 |
| | (1.45) | (3.00) | (2.06) | (2.91) | (3.25) | (4.24) | (6.00) | (6.71) |

The values in this table represent the conditional VaR quantiles for different confidence intervals, for example $r_p95$ is the 5% level. The estimates use Hill estimators based on the Huisman et al. (2001) procedure from the AR(1)-GARCH(1, 1) filtered returns. The blocks of returns used are two days n = 2, four days n = 4 and five days (weekly) n = 5. Conditional estimates from fitting a GARCH (1, 1) model with normal innovations are in parentheses. Values are expressed in percentages.